\documentclass[aps,prl,twocolumn,superscriptaddress,floatfix]{revtex4-2}

\usepackage{graphicx}
\usepackage{dcolumn}
\usepackage{bm}
\usepackage{amsmath,amssymb}
\usepackage{siunitx}
\usepackage{xcolor}
\usepackage[colorlinks=true,citecolor=blue,linkcolor=blue,urlcolor=blue]{hyperref}
\usepackage{orcidlink}

\begin{document}

\preprint{JLAB-ACC-26-4924}
\title{Cryogenic Enhancement of Electron Spin Polarization from a Strained GaAs/GaAsP Superlattice Photocathode}

\author{Matt Grau\,\orcidlink{0000-0002-2684-6923}}
\email{mgrau@odu.edu}
\affiliation{Department of Physics, Old Dominion University, Norfolk, Virginia 23529, USA}

\author{Colin Kirk} %
\affiliation{Department of Electrical \& Computer Engineering, Old Dominion University, Norfolk, Virginia 23529, USA}

\author{Greg Blume\,\orcidlink{0000-0003-4223-2259}}
\affiliation{Department of Physics, Old Dominion University, Norfolk, Virginia 23529, USA}

\author{John Hill\,\orcidlink{0009-0002-7581-8184}}
\affiliation{Department of Electrical \& Computer Engineering, Old Dominion University, Norfolk, Virginia 23529, USA}

\author{Sushil Poudel\,\orcidlink{0009-0009-8152-7093}}
\affiliation{Department of Electrical \& Computer Engineering, Old Dominion University, Norfolk, Virginia 23529, USA}

\author{Alimohammed Kachwala\,\orcidlink{0000-0003-0175-6478}}
\affiliation{Thomas Jefferson National Accelerator Facility, Newport News, Virginia 23606, USA}

\author{Marcy Stutzman\,\orcidlink{0000-0001-7460-4796}}
\affiliation{Thomas Jefferson National Accelerator Facility, Newport News, Virginia 23606, USA}

\author{Joseph Michael Grames\,\orcidlink{0000-0001-9390-8752}}
\affiliation{Thomas Jefferson National Accelerator Facility, Newport News, Virginia 23606, USA}

\author{Sylvain Marsillac\,\orcidlink{0000-0003-0826-8119}}
\affiliation{Department of Electrical \& Computer Engineering, Old Dominion University, Norfolk, Virginia 23529, USA}

\author{Matt Poelker\,\orcidlink{0000-0002-5441-137X}}
\affiliation{Thomas Jefferson National Accelerator Facility, Newport News, Virginia 23606, USA}

\date{\today}

\begin{abstract}
We report electron spin polarization of $95.0 \pm 0.8\,\mathrm{(stat)} \pm 2.4\,\mathrm{(sys)}\%$ from a strained GaAs/GaAsP superlattice photocathode grown by metalorganic chemical vapor deposition (MOCVD) and cooled to dry-ice temperature (\SI{195}{K}). We achieved this polarization with 97.8\% circularly polarized excitation light and a quantum efficiency of 0.7\% at the peak polarization wavelength.
This measurement exceeds the values of previously reported GaAs-based photocathode polarizations, which have clustered near 92\% for two decades.
We vary the temperature of the cathode and measure the polarization and quantum-efficiency spectra at \SI{295}{K}, \SI{273}{K}, \SI{195}{K}, and \SI{77}{K}.
The polarization rises from $91.2(1)\%$ at \SI{295}{K} to its maximum at \SI{195}{K}, while the spectral peak shifts from \SI{775}{nm} to \SI{739}{nm} (\SI{78}{meV}) over the full temperature range, tracking the widening band gap.
The spectra probe two depolarization mechanisms: a thermalized transport channel that is suppressed on cooling, and energy-dependent hot-electron relaxation that persists and sets the low-temperature saturation.
The polarization recovers after a full cooling and warmup cycle and is stable while the quantum efficiency decays, which disfavors surface energy filtering as the origin of the gain.
These results indicate that modest cooling to \SI{195}{K}, for which dry ice suffices, is a practical route to higher-polarization GaAs-based electron sources.
\end{abstract}

\maketitle

\textit{Introduction.---}High polarization electron beams are central to parity-violating electron scattering~\cite{Aulenbacher1997,Sinclair2007,Adderley2010,Grames2011,Adderley2023}, to measurements of nucleon spin structure~\cite{Kuhn2009} and elastic form factors by polarization transfer~\cite{Jones2000,Perdrisat2007}, to polarized-positron production~\cite{Abbott2016}, and to future collider experiments, notably the Electron-Ion Collider (EIC)~\cite{Accardi2016,AbdulKhalek2022,Wang2022,Litvinenko2026}, where the statistical reach at fixed beam current scales quadratically in the beam polarization.
GaAs-based photocathodes activated to negative electron affinity (NEA) are the standard for high-current high-polarization electron sources~\cite{Pierce1975,Pierce1976}. Strained GaAs epilayers lift the degeneracy between heavy-hole (HH) and light-hole (LH) valence bands that limits bulk GaAs to 50\% polarization~\cite{Maruyama1991,Nakanishi1991}, and strained superlattices (SLs) combine strain with quantum confinement to push the emitted polarization toward unity~\cite{Omori1991,Maruyama2004,Subashiev2004}.
The measured spin polarization from strained-layer and strained-superlattice cathodes has remained near 90--92\% for two decades: $92\pm6\%$~\cite{Nishitani2005}, 91--92\%~\cite{Gerchikov2006,Mamaev2008}, and 92\% with 1.6\% quantum efficiency (QE)~\cite{Jin2013}.
Over the same period, QE and operational lifetime have improved substantially~\cite{Liu2016,Biswas2023,Belfore2023,Masters2026,Wang2024,Litvinenko2026}.

The gap between measured polarization and the theoretical limit comes from depolarization of the electrons during the excitation, transport, and emission processes. These mechanisms all depend strongly on temperature: thermally broadened LH leakage into the HH excitation window~\cite{Subashiev2004,Varshni1967}, spin relaxation of thermalized electrons during transport to the NEA surface, dominated by exchange scattering from holes in heavily p-doped material~\cite{Fishman1977,Zerrouati1988,Liu2017,Ohki2017}, and spin relaxation during the hot-electron cascade~\cite{Chubenko2021,Callahan2025}.
For strained-superlattice photocathodes, changing temperature lets us probe these depolarization channels and distinguish polarization gain in the bulk or during transport from surface energy-filtering artifacts~\cite{Levenson2025}.
Cooling bulk GaAs to \SI{77}{K} is known to recover its full 50\% selection-rule value~\cite{Liu2017}, but low-temperature photoemission studies have otherwise been confined to bulk GaAs or surface-physics contexts~\cite{Pierce1976,Allenspach1984}, and to our knowledge no temperature-dependent measurement of the emitted-electron polarization of a strained-superlattice photocathode has been reported.

\begin{figure}[tb]
    \centering
    \includegraphics[width=\columnwidth]{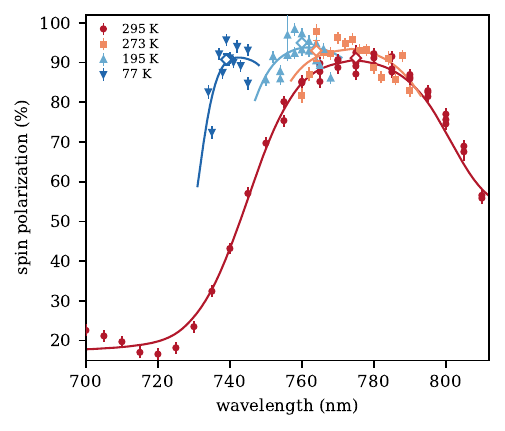}
    \caption{Delivered beam polarization versus excitation wavelength at \SI{295}{K}, \SI{273}{K}, \SI{195}{K}, and \SI{77}{K} (filled points; error bars are the counting error and a 1.5-percentage-point scan-to-scan repeatability uncertainty, added in quadrature), with the empirical description of the spectra multiplied by the measured laser circular polarization (solid curves). The model uses the peak-shift-inferred cathode temperatures, the cold-saturation transport ceiling, thermal edge smearing, and the LO-occupation onset law described in the text. Open diamonds mark the quoted peak values, the error-weighted means of all production-spectrum measurements within \SI{3}{nm} of each peak. A 2.49\% relative Sherman-function uncertainty applies to all points; the separate laser-calibration uncertainty applies only to helicity-normalized quantities and is not shown.}
    \label{fig:pol-vs-wavelength}
\end{figure}

Here we use temperature to explore and suppress electron spin depolarization in an MOCVD-grown strained GaAs/GaAsP superlattice, measuring the spin-polarization spectra shown in Fig.~\ref{fig:pol-vs-wavelength} and QE spectra at nominal cathode temperatures of \SI{295}{K}, \SI{273}{K}, \SI{195}{K}, and \SI{77}{K}.
The peak polarization rises from $91.2(1)\%$ at room temperature to $95.0\pm0.8\,\mathrm{(stat)}\pm2.4\,\mathrm{(sys)}\%$ at \SI{195}{K}, while the QE at the polarization peak remains near 1\%. Correcting for the measured circular polarization of the excitation gives helicity-normalized polarizations of $97.1(0.8)\%$ at \SI{195}{K} and $94.3(1.3)\%$ at \SI{77}{K}; their respective laser-calibration uncertainties are 0.3 and 1.2 percentage points, in addition to Sherman-function uncertainties of 2.4 and 2.3 percentage points.
Even the bare measured value exceeds the central values of the prior benchmark measurements, which cluster near 92\%~\cite{Nishitani2005,Mamaev2008,Jin2013}, and the largest delivered value is reached at dry-ice temperature.
The temperature dependence of the peak wavelength tests the calculated band-structure shift, the temperature dependence of the peak polarization constrains an effective transport-depolarization contribution, and stability and reversibility measurements of the QE and polarization test for surface energy-filtering artifacts, in which an evolving activation layer raises the measured polarization without improving the source material~\cite{Levenson2025}.

\textit{Experiment.---}The photocathode is a strained GaAs/GaAsP superlattice grown by MOCVD following Refs.~\cite{Belfore2023,Masters2026}.
The superlattice comprises 30 pairs of \SI{3.8}{nm} strained GaAs wells and \SI{2.8}{nm} GaAs$_{0.65}$P$_{0.35}$ barriers, Zn doped at a concentration of \SI{5e17}{cm^{-3}}. All layer thicknesses are growth targets. The superlattice sits inside an optical cavity between a \SI{5}{nm} GaAs surface layer carbon-doped at \SI{5e19}{cm^{-3}} and a distributed Bragg reflector (DBR) formed by 12 pairs of GaAs$_{0.65}$P$_{0.35}$ and In$_{0.30}$Al$_{0.70}$P layers. A \SI{338}{nm} GaAs$_{0.65}$P$_{0.35}$ spacer separates the superlattice from the DBR. The p-GaAs substrate, metamorphic buffer, DBR, and spacer are Zn doped at \SI{5e18}{cm^{-3}}. Reflectance measurements at normal incidence confirm the DBR stopband at the operating wavelengths.
We initially heat cleaned the sample at \SI{550}{\celsius} for \SI{1.5}{h} to remove surface oxides. Before each subsequent test we heated it at \SI{350}{\celsius} to reset the NEA condition, then reactivated it with cesium and NF$_3$. Each activation required 6--10 Cs/NF$_3$ yo-yo cycles and ended when an additional cycle increased the photocurrent by less than 5\%. During measurements, the chamber pressure was approximately \SIrange{3e-12}{1e-11}{Torr}; during activation it rose as high as \SI{1e-9}{Torr}.

The sample holder is a hollow stainless-steel reservoir terminated by a thin molybdenum end plate. We mounted the GaAs photocathode against the end plate with indium foil and retained it with a tantalum cup, leaving a circular active area \SI{12.9}{mm} in diameter. We filled the reservoir successively with ice water, dry ice, and liquid nitrogen~\cite{Liu2017}. Direct cathode thermometry was not available, so we report the nominal coolant temperatures of \SI{273}{K}, \SI{195}{K}, and \SI{77}{K}.
We measured the polarization and quantum efficiency spectra with excitation light from a broadly tunable supercontinuum light source (NKT Photonics SuperK) covering a wavelength range of \SIrange{600}{810}{nm}. We prepared circularly polarized light using a linear polarizer and a zero-order \SI{780}{nm} quarter-wave plate, and reversed the helicity by inserting a half-wave plate during measurement to cancel instrumental asymmetries. We characterized the wavelength-dependent circular polarization of the excitation light at the cathode to be 89--100\% over the polarization scans.
We report two quantities: the electron spin polarization delivered to the polarimeter $P_{\mathrm{beam}}(\lambda)$, measured with the actual circular polarization of the excitation, and the polarization corrected for the laser circular polarization, $P_{\mathrm{corr}}(\lambda)=P_{\mathrm{beam}}(\lambda)/C(\lambda)$, where $C$ is the measured circular polarization of the light at the cathode.
The measurement of $C(\lambda)$ used a rotating zero-order \SI{780}{nm} half-wave plate and a fixed Glan polarizer, calibrated for the plate's wavelength-dependent retardance (SM).
The corrected value is the quantity we use to compare depolarization across temperatures and with prior work.
We obtained the QE from the photocurrent and the incident power at each wavelength.
We measured polarization with a retarding-field micro-Mott polarimeter~\cite{Gay1992,McCarter2010} operated at \SI{20}{kV} with effective Sherman function $S_{\mathrm{eff}}=0.201(5)$, calibrated by comparison with the CEBAF \SI{5}{MeV} Mott polarimeter~\cite{Steigerwald2001,McCarter2010,Masters2026}. The Supplemental Material (SM)~\cite{SupplementalMaterial} gives the full layer structure and details the polarimeter operation and calibration.
We quote the statistical uncertainty of the helicity-pair asymmetry on each polarization point; where repeated measurements within the peak window scatter beyond counting statistics, we inflate the uncertainty by $\sqrt{\chi^2_\nu}$.
We quote peak polarizations as the weighted mean of all measurements within \SI{3}{nm} of the peak wavelength; the $\sqrt{\chi^2_\nu}$ inflation of the peak uncertainty reaches a factor of 11 at \SI{77}{K}, where successive scans scatter by several percentage points.
For helicity-normalized values we propagate the local uncertainty of $C(\lambda)$ as a separate laser-calibration component. We also add a 2.49\% relative Sherman-function uncertainty, common to all polarization points. We exclude both correlated calibration components from the spectral-fit weights and include them when comparing absolute peak polarizations with published benchmarks.

A high measured polarization can also be a surface artifact. As the activation layer degrades, the surface can shift from negative to positive electron affinity, which filters the emitted electrons in energy and raises the measured polarization of bulk GaAs without any change in the material itself~\cite{Levenson2025}. Two controls address this concern.
First, the polarization is reversible. After the full cooling sequence and warmup, we recovered the room-temperature peak polarization within uncertainty ($+1.1(2.1)$ percentage points relative to the pre-cooling value); three same-day \SI{775}{nm} checks span 89.1--92.5\%, with the largest difference of $+3.4(2.4)$ percentage points (counting plus repeatability uncertainty).
Second, the polarization does not track activation decay. At \SI{77}{K} the QE at the operating wavelength decreased by 43\% over the measurement day, while the polarization measurements showed no systematic increase. A repeat dry-ice cooldown late in the campaign, after substantial QE decay, gave a \emph{lower} helicity-normalized peak polarization ($90.1(1.1)\%$), the opposite of the energy-filtering signature.
These observations strongly disfavor activation degradation or energy filtering as the origin of the high-polarization state; they do not imply that the activation layer is unchanged, but they show that the polarization gain is not correlated with activation aging.

\begin{figure}[tb]
    \centering
    \includegraphics[width=\columnwidth]{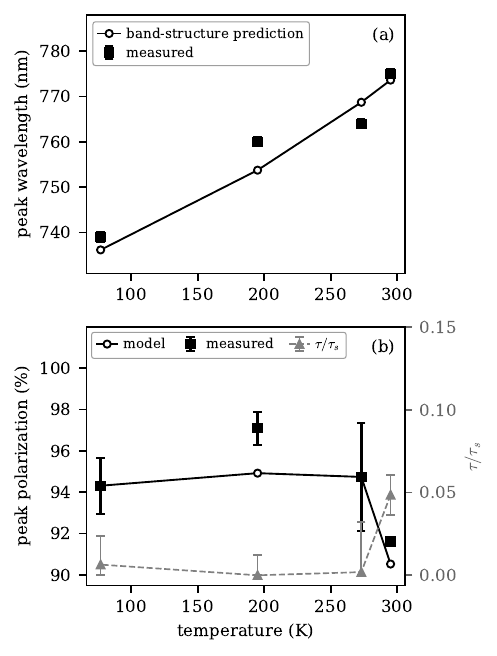}
    \caption{(a)~Measured polarization-peak wavelength versus temperature (squares) with the band-structure prediction (circles). (b)~Measured helicity-normalized peak polarization (squares) with the model (circles) and the effective transport factor $\tau/\tau_s$ extracted under the temperature-independent-background assumption (right axis, triangles). This factor is consistent with zero at and below \SI{273}{K}.}
    \label{fig:model}
\end{figure}

\textit{Polarization spectra.---}At \SI{295}{K} the spectrum has the form characteristic of high-quality strained SLs~\cite{Maruyama2004,Nishitani2005,Jin2013}, with a delivered peak polarization of $91.2(1)\%$ at \SI{775}{nm}, or $91.6(1)\%$ when corrected for the 99.5\% circular polarization at that wavelength.
Cooling the sample produces three correlated changes.
The delivered peak polarization rises to $93.0(2.6)\%$ at \SI{273}{K} and $95.0(0.8)\%$ at \SI{195}{K}, corresponding to helicity-normalized values of $94.7(2.6)\%$ and $97.1(0.8)\%$ with $C=98.1\%$ and 97.8\% circular polarization at the respective peaks.
At \SI{77}{K} the delivered peak of $90.8(1.3)\%$ is lower largely because the laser circular polarization at the \SI{739}{nm} peak drops to 96.3\%; the helicity-normalized value, $94.3(1.3)\%$, has a lower central value but remains consistent with the \SI{195}{K} result within the combined statistical and laser-calibration uncertainties (a difference of 1.4 standard deviations). The data therefore do not resolve a further polarization increase below \SI{195}{K}.
The quoted peaks use all production-spectrum measurements within \SI{3}{nm} of each peak, and their uncertainties include the observed point-to-point scatter. The highest individual scan points at \SI{273}{K}, \SI{195}{K}, and \SI{77}{K} reach helicity-normalized values statistically consistent with complete spin polarization.
The absence of a resolved further increase is consistent with the dominant temperature-dependent depolarization channel being strongly suppressed already at dry-ice temperature.
The peak polarization wavelength shifts from \SI{775}{nm} at \SI{295}{K} to \SI{739}{nm} at \SI{77}{K}, a shift of \SI{78}{meV}, close to the \SI{83}{meV} Varshni band-gap shift expected at fixed confinement offset for the nominal coolant temperatures~\cite{Varshni1967,Vurgaftman2001} [Fig.~\ref{fig:model}(a)]. Since we do not measure the emitting-surface temperature directly, this comparison is a consistency check on band-edge shift rather than an independent thermometric calibration.
The spectral peak also narrows on cooling. At \SI{77}{K} the scans cover \SIrange{734}{745}{nm}, and we infer the peak width there from the phenomenological description of the narrowing (see below).

The maximum helicity-normalized polarization of $97.1\pm0.8\,\mathrm{(stat)}\pm0.3\,\mathrm{(laser)}\pm2.4\,\mathrm{(Sherman)}\%$ is, to our knowledge, the highest central value reported from a GaAs-based photocathode, exceeding the previous benchmarks, $92\pm6\%$~\cite{Nishitani2005} and 92\%~\cite{Mamaev2008,Jin2013}, by more than five percentage points, and even the delivered beam polarization, $95.0(0.8)\%$, exceeds these values. The SM tabulates a comparison with published strained-layer and superlattice results.

\begin{figure}[tb]
    \centering
    \includegraphics[width=\columnwidth]{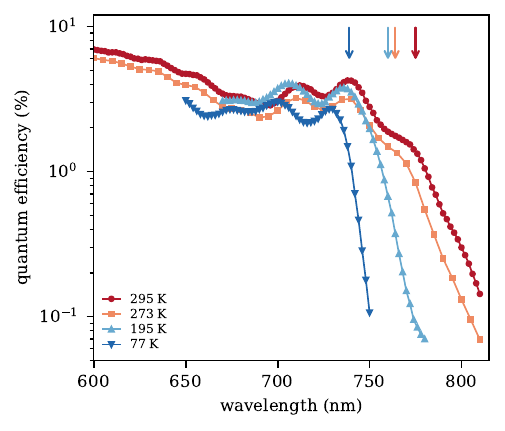}
    \caption{Quantum efficiency versus excitation wavelength at each temperature, from the scans taken closest in time to the polarization runs. Arrows mark the operating (polarization-peak) wavelengths.
     }
    \label{fig:qe-vs-wavelength}
\end{figure}

\textit{Quantum efficiency.---}
We define the nominal operating wavelength at each temperature as the polarization-peak wavelength.
The QE spectra (Fig.~\ref{fig:qe-vs-wavelength}) shift with the band edge. At the fixed room-temperature operating wavelength of \SI{775}{nm}, the QE drops by a factor of 1.6 at \SI{273}{K} and by more than an order of magnitude at \SI{195}{K} as the absorption edge moves through it~\cite{Liu2017}. At the operating wavelength of each temperature, the QE is 1.4\%, 1.4\%, 0.7\%, and 1.3\% at \SI{295}{K}, \SI{273}{K}, \SI{195}{K}, and \SI{77}{K}.
Polarization and QE enter the source performance differently. The polarization of a beam is fixed at the point of photoemission, whereas the intensity at $\sim$1\% QE can be raised with illumination power, within laser and cathode heat-load and lifetime limits.
Cooling therefore improves the quantity that is fixed at the source, while the measured QE of $\sim$1\% remains adequate for typical operation.
For experiments limited instead by laser power or by the surface charge limit at high bunch charge~\cite{Togawa1998,Clendenin2005}, the composite metric $P^{2}\,\mathrm{QE}$ at the operating wavelength remains within a factor of two of its room-temperature value across the range (0.6--1.2\%).

\textit{Model.---}We model the spectra in two stages, with the adjustable spectral shape constrained by the room-temperature data.
The model extends to lower temperatures through the known temperature dependence of the band gaps, plus one fitted transport factor per temperature.
The first stage is a band-structure calculation with no adjustable parameters. We diagonalize a four-band Luttinger-Kohn Hamiltonian with Bir-Pikus strain for the superlattice valence minibands and a nonparabolic conduction miniband, using the nominal layer structure and literature material constants, and apply the $\sigma^{+}$ dipole selection rules~\cite{Vurgaftman2001,Subashiev2004}.
This calculation fixes the heavy-hole miniband edge \SI{133}{meV} above the bulk GaAs gap, the HH--LH splitting $\delta=\SI{105}{meV}$ (with \SI{97}{meV} arising from strain and the rest from confinement), and the decay of the excitation polarization above the edge through $k_{\parallel}$ valence-band mixing~\cite{Subashiev2004}.
The Varshni gap laws set the temperature shift of every edge~\cite{Varshni1967,Vurgaftman2001}.
Because the material parameters derive largely from low-temperature measurements, these quantities provide a reliable temperature-dependent baseline over the measured range.
Next we assemble the transitions into a spectrum,
\begin{equation}
P(\hbar\omega,T)=\frac{1}{1+\tau/\tau_s}\,
\frac{\sum_i P_i\,s_i\,a_i\,g_i(\hbar\omega)+P_{\mathrm{bg}}\,a_{\mathrm{bg}}}{\sum_i a_i\,g_i(\hbar\omega)+a_{\mathrm{bg}}},
\label{eq:model-eq}
\end{equation}
Each term carries one physical ingredient.
Transition $i$ has absorption weight $a_i$, including a two-dimensional Coulomb enhancement, and $\sigma^{+}$ polarization $P_i$, both taken from the band-structure calculation.
The line shape $g_i$ is a Gaussian of width $\Gamma$ centered on the calculated transition energy $E_i$ shifted by a fitted rigid offset $\Delta E_0$, with $\Gamma$ and $\Delta E_0$ common to all transitions.
The factor $s_i=\exp[-(\Delta E_{c,i}/E_\kappa)^4]$ is the spin survival of the hot-electron cascade, where $\Delta E_{c,i}$ is the conduction excess energy of transition $i$ and $E_\kappa$ is the excess energy beyond which the cascade erases the spin memory; the quartic form is the phonon-cascade integral of the D'yakonov--Perel' rate~\cite{Dyakonov1972,Chubenko2021,Callahan2025}.
The term $a_{\mathrm{bg}}$, carrying polarization $P_{\mathrm{bg}}$, is an effective weakly polarized contribution whose microscopic origin is not fixed here (SM).
The prefactor $1/(1+\tau/\tau_s)$ describes spin relaxation of thermalized electrons during transport, with $\tau$ the emission time and $1/\tau_s$ the spin-relaxation rate~\cite{Fishman1977,Liu2017}.
We fit the room-temperature spectrum and the three lower-temperature peak polarizations with six shared spectral parameters and $\tau/\tau_s$ at each temperature: the edge shift $\Delta E_0$, which absorbs the band-offset and layer-thickness uncertainty, the width $\Gamma$, the cascade scale $E_\kappa$, the exciton strength, and the background amplitude and polarization.
The room-temperature spectrum fixes the spectral shape, and each lower temperature contributes only its measured peak polarization, which determines its $\tau/\tau_s$ (SM).

We take the three changes of the spectra upon cooling in turn: the shift of the peak wavelength, the rise of the peak polarization, and the narrowing of the peak.
With the room-temperature edge shift fixed, the peak wavelength contains no further fitted temperature dependence, and the predicted peaks lie within \SI{6}{nm} of the measured values over the full \SI{36}{nm} shift [Fig.~\ref{fig:model}(a)].
If the effective background is temperature independent, the fit attributes the polarization gain to suppression of the transport factor, which falls from $\tau/\tau_s=0.049(12)$ at \SI{295}{K} to values consistent with zero at and below \SI{273}{K} [Fig.~\ref{fig:model}(b)].
The modeled gain of 4.4 percentage points is close to the measured 5.5; within this parameterization, the absence of a resolved further increase below \SI{195}{K} follows because the residual deficit at the peak (background, band mixing, and cascade depolarization) is held temperature independent.
Three mechanisms govern conduction-electron spin relaxation in p-type III--V semiconductors~\cite{OpticalOrientation1984,Zutic2004,Wu2010}: D'yakonov--Perel' (DP) precession about the momentum-dependent effective field of the inversion-asymmetric lattice~\cite{Dyakonov1972}, Bir--Aronov--Pikus (BAP) electron--hole exchange scattering~\cite{Fishman1977,Zerrouati1988}, and a small Elliott--Yafet contribution at our acceptor density of \SI{5e17}{cm^{-3}}~\cite{Chubenko2021}.
DP dominates at room temperature and BAP becomes the larger thermalized channel on cooling.
The transport factor we extract, however, falls to a value consistent with zero already at \SI{273}{K}, a change far larger than any thermalized bulk rate produces over a \SI{22}{K} interval, and may reflect temperature dependence of the emission time, energy relaxation, additional transport channels, or the low-polarization contribution represented by $a_{\mathrm{bg}}$~\cite{Ohki2017}.
We describe the peak narrowing below room temperature phenomenologically.
A simplified two-edge description of the spectra (SM) locates the narrowing in the blue-side depolarization onset, which falls from \SI{92(8)}{meV} above the heavy-hole edge at \SI{295}{K}, close to the fixed HH--LH splitting, to \SI{48(8)}{meV} at the \SI{77}{K} coolant setting.
The onset can be parameterized as $\delta_{\mathrm{eff}}(T)\simeq\hbar\omega_{\mathrm{LO}}[1+4.2(5)\,\bar{n}_{\mathrm{LO}}]$, where $\bar{n}_{\mathrm{LO}}$ is the Bose occupation of the \SI{36.7}{meV} LO phonon evaluated at the peak-shift-inferred cathode temperature. The onset approaches the single LO-phonon energy as the phonons freeze out.
This correlation is consistent with LO-phonon freeze-out contributing to the narrowing, but the relation is purely empirical and does not establish a microscopic mechanism or a unique cause.

\textit{Conclusion.---}Our temperature-dependent polarization and QE spectra of a strained GaAs/GaAsP superlattice photocathode distinguish an effective transport contribution from energy-dependent spin relaxation during the hot-electron cascade.
The increase in polarization on cooling is consistent with suppressed transport depolarization, raising the delivered peak polarization from $91.2(1)\%$ at \SI{295}{K} to $95.0\pm0.8\,\mathrm{(stat)}\pm2.4\,\mathrm{(sys)}\%$ at \SI{195}{K}, corresponding to a helicity-normalized polarization of $97.1\pm0.8\,\mathrm{(stat)}\pm0.3\,\mathrm{(laser)}\pm2.4\,\mathrm{(Sherman)}\%$.
Within the temperature-independent-background parameterization, the absence of a resolved further increase on cooling to \SI{77}{K} is consistent with the surviving energy-dependent hot-electron contribution, and the QE remains near 1\% at the polarization-optimized wavelength throughout.
The peak wavelength follows the band-structure prediction within \SI{6}{nm} at every temperature.
Reversibility of the QE and polarization spectra, and a repeat cooldown in which activation decay lowered rather than raised the polarization, strongly disfavor surface-filtering artifacts.
The peak narrowing can be parameterized by the LO-phonon occupation, consistent with a possible role for LO-phonon freeze-out.
The gain is consistent with suppression of bulk or transport depolarization, and the full gain is available at dry-ice temperature, compatible with dc high-voltage photogun operation.
Combined with high-QE superlattice designs~\cite{Liu2016,Biswas2023,Masters2026}, cooled operation provides next-generation experiments~\cite{Wang2022,Litvinenko2026} with polarized electron sources at the highest central polarization yet reported for GaAs photocathodes.

\begin{acknowledgments}
This material is based upon work supported by the U.S.\ Department of Energy, Office of Science, Office of Nuclear Physics under Contract No.\ 89243126CSC000213 and Award No.\ DE-SC0025519.
\end{acknowledgments}

\nocite{Belfore2022,Rusetsky2022,Vurgaftman2021}
\bibliography{references}

\end{document}

\title{Supplemental Material for ``Cryogenic Enhancement of Electron Spin Polarization from a Strained GaAs/GaAsP Superlattice Photocathode''}
\date{\today}
\maketitle

\section{Apparatus}

We mounted the GaAs photocathode against the thin molybdenum end plate of a hollow stainless-steel reservoir with indium foil and retained it with a tantalum cup, leaving a circular active area \SI{12.9}{mm} in diameter. We filled the reservoir successively with ice water, dry ice, and liquid nitrogen, following Ref.~\cite{Liu2017}.
Direct cathode thermometry was not available. We therefore identify the cold conditions by the nominal equilibrium temperatures of the reservoir coolants, \SI{273}{K}, \SI{195}{K}, and \SI{77}{K}; the band-edge shift provides a consistency check on those assignments, not an independent temperature measurement of the emitting surface.
We measured all quoted QE spectra at a single fixed location on the sample, near the center of the illuminated area, and we kept that location for the polarization measurements; for this sample the maximum-QE and maximum-transmission spots nearly coincide, so the choice does not affect the quoted values.

\section{Sample structure and activation}
\begin{figure}[htb]
    \centering
    \includegraphics[width=\columnwidth]{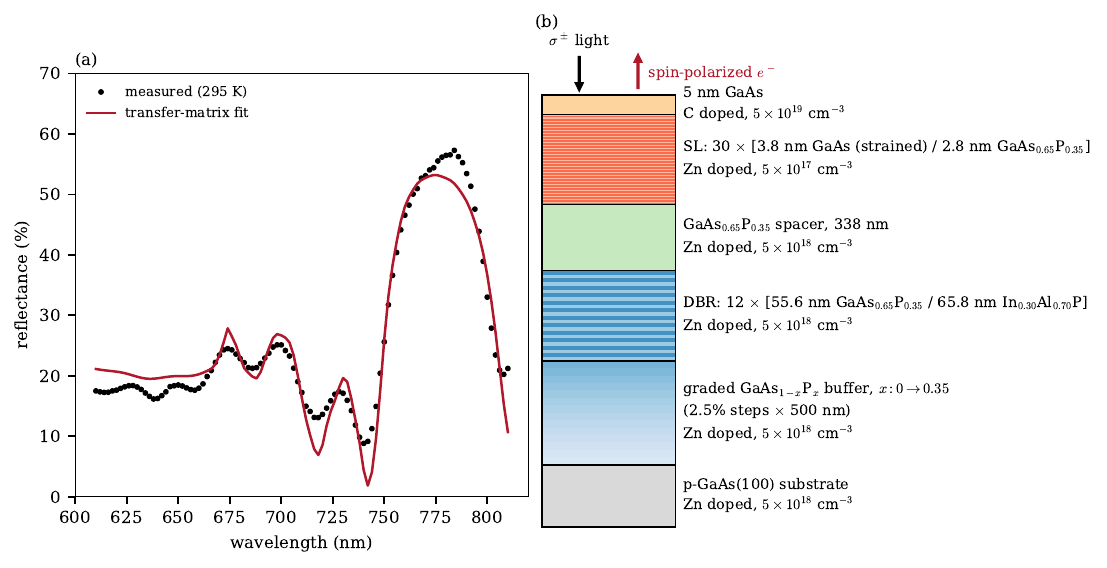}
    \caption{(a)~Measured normal-incidence reflectance of the tested piece at \SI{295}{K} with the transfer-matrix calculation of the stack at fitted thicknesses; the calculated reflectance is scaled and offset to match the data (see text). (b)~Layer structure of photocathode sample 26R016 (not to scale). Illumination and photoemission occur through the \SI{5}{nm} GaAs surface layer, and the \SI{338}{nm} spacer and the superlattice form the optical cavity between the DBR and the surface layer.}
    \label{fig:sm-stack}
\end{figure}

We grew the sample used in this study, which we refer to as 26R016, by MOCVD with the same process and precursors as Ref.~\cite{Belfore2023}, with more superlattice pairs (30 versus 14) and an adjusted optical cavity.
The structure begins on a p-GaAs(100) on-axis substrate (Zn, \SI{5e18}{cm^{-3}}).
All layer thicknesses below are growth targets.
A compositionally step-graded metamorphic GaAs$_{1-x}$P$_x$ buffer brings the lattice constant from GaAs to GaAs$_{0.65}$P$_{0.35}$ in 2.5\% composition steps of \SI{500}{nm} each, ending in a GaAs$_{0.625}$P$_{0.375}$ overshoot layer and a \SI{2.5}{\micro\meter} GaAs$_{0.65}$P$_{0.35}$ fallback layer.
On this relaxed template sit a distributed Bragg reflector (DBR) formed by 12 pairs of \SI{55.6}{nm} GaAs$_{0.65}$P$_{0.35}$ and \SI{65.8}{nm} In$_{0.30}$Al$_{0.70}$P, a \SI{338}{nm} GaAs$_{0.65}$P$_{0.35}$ spacer, and the active region, 30 periods of \SI{3.8}{nm} compressively strained GaAs wells and \SI{2.8}{nm} unstrained GaAs$_{0.65}$P$_{0.35}$ barriers.
A \SI{5}{nm} GaAs surface layer, carbon doped at \SI{5e19}{cm^{-3}}, terminates the structure for activation.
The substrate, metamorphic graded buffer, DBR, and spacer are Zn doped at \SI{5e18}{cm^{-3}}, the superlattice at \SI{5e17}{cm^{-3}}, and the surface layer at \SI{5e19}{cm^{-3}}.
Figure~\ref{fig:sm-stack}(b) shows the stack.

Figure~\ref{fig:sm-stack}(a) compares the measured normal-incidence reflectance of the tested piece with a transfer-matrix calculation of this structure using optical constants from Ref.~\cite{Vurgaftman2021}.
At room temperature the off-stopband reflectance is 16--25\% over \SIrange{610}{700}{nm}, and the stopband reaches 57\%; at \SI{77}{K} the corresponding values are 15--24\% and 59\%.
The cavity-reflection minimum blueshifts by \SI{6}{nm}, from \SI{740}{nm} to \SI{734}{nm}, on cooling to \SI{77}{K}.
The minimum corresponds to the fifth-order half-wave resonance of the cavity that the spacer and superlattice form between the DBR and the surface layer.
A transfer-matrix fit that floats the DBR and spacer thicknesses, with the calculated reflectance scaled and offset to match the data, reproduces the spectrum and returns DBR thicknesses within 0.4\% of the design values and the spacer at \SI{327(1)}{nm}, 3\% below design; the reduced specular amplitude is consistent with scatter from the cross-hatched surface of the metamorphic stack.

We first heat cleaned the sample at \SI{550}{\celsius} for \SI{1.5}{h} to remove surface oxides and activated it to negative electron affinity with cesium and NF$_3$, using the same yo-yo procedure as previous JLab and ODU activations of this cathode family~\cite{Belfore2022,Masters2026}.
Before each subsequent test we heat cleaned the sample at \SI{350}{\celsius} to reset the NEA condition, reactivated it, and recorded a room-temperature baseline polarization spectrum on the fresh activation before cooling.
A re-activation preceded each of the dry-ice-day and ice-water-day measurements.
Each activation required 6--10 Cs/NF$_3$ cycles and ended when an additional cycle increased the photocurrent by less than 5\%~\cite{Masters2026}.

\section{Mott polarimetry and uncertainty budget}
During polarization measurements, the chamber pressure ranged from approximately \SI{3e-12}{Torr} to \SI{1e-11}{Torr}; during activation it reached as high as \SI{1e-9}{Torr}.
We operated the retarding-field micro-Mott polarimeter~\cite{Gay1992,McCarter2010} at \SI{20}{kV} with a gold-plated copper target.
We formed the scattering asymmetry from helicity pairs as $A=(\sqrt{N^+_L N^-_R}-\sqrt{N^-_L N^+_R})/(\sqrt{N^+_L N^-_R}+\sqrt{N^-_L N^+_R})$, which cancels detector-efficiency and beam-intensity asymmetries to first order.
The effective Sherman function, $S_{\mathrm{eff}}=0.201(5)$, was calibrated by measuring the same superlattice photocathode material with this instrument and the CEBAF \SI{5}{MeV} Mott polarimeter~\cite{Steigerwald2001,McCarter2010}.

We take the peak wavelength as the scan wavelength with the highest error-weighted mean polarization, quote peak polarizations as the error-weighted mean of all measurements within \SI{3}{nm} of it, and inflate the uncertainty by $\sqrt{\chi^2_\nu}$ when repeated measurements within the window scatter beyond counting statistics.

\emph{Laser-polarization correction.}
We measured the circular fraction $C(\lambda)$ of the excitation light at the cathode with a rotating zero-order \SI{780}{nm} half-wave plate followed by a fixed Glan polarizer, calibrated for the wavelength-dependent retardance of the plate~\footnote{Thorlabs, ``780-nm wave-plate retardance data,'' \url{https://media.thorlabs.com/contentassets/93eeaf81d3884ffe9b46fc6f6341e0a8/780_wpdata_xls.xls?v=1202115331} (accessed August 11, 2026).}.
We repeated the measurement in two independent optical trials, averaged them at each characterized wavelength, and interpolated between wavelengths with a shape-preserving spline.
At the four quoted peak wavelengths, \SI{775}{nm}, \SI{764}{nm}, \SI{760}{nm}, and \SI{739}{nm}, the circular fractions are $0.9949\pm0.0014$, $0.9810\pm0.0048$, $0.9780\pm0.0029$, and $0.9629\pm0.0119$, where the uncertainty is the standard error of the two trials.
Because the retardance data are the manufacturer's theoretical curve, this uncertainty describes the repeatability of the two optical trials and not a possible unit-specific deviation from that curve.

The helicity-normalized polarization follows as $P_{\mathrm{corr}}=P_{\mathrm{beam}}/C$, and the calibration uncertainty propagates as $\sigma_{P,C}=P_{\mathrm{corr}}\sigma_C/C$.
Because the same calibration curve is shared by repeated spectra, we retain this component separately rather than treating it as independent point noise in the spectral fits; the 2.49\% Sherman-function uncertainty is also retained separately.

The resulting helicity-normalized peak polarizations at \SI{295}{K}, \SI{273}{K}, \SI{195}{K}, and \SI{77}{K} are $91.6\pm0.1\pm0.1\pm2.3\%$, $94.7\pm2.6\pm0.4\pm2.4\%$, $97.1\pm0.8\pm0.3\pm2.4\%$, and $94.3\pm1.3\pm1.2\pm2.3\%$, respectively, where the uncertainties are statistical plus scan-to-scan scatter, laser calibration, and the Sherman function.
At \SI{77}{K} the six window points span $87.5$--$95.7\%$ (delivered) with individual statistical errors of $0.2$--$0.5\%$, giving $\chi^2_\nu=115$ and an inflation factor of 11.
The \SI{273}{K} spectrum accumulated overnight on a single ice-water fill, topped off immediately before the scan, so its window scatter also includes slow drift of the activation and coolant over several hours.
We do not quote single-scan maxima, because selecting the highest of several mutually inconsistent repeats biases the estimate upward.
The highest single-scan helicity-normalized values are $99.9\pm0.8\,\mathrm{(stat)}\pm0.5\,\mathrm{(laser)}\pm2.5\,\mathrm{(Sherman)}\%$ at \SI{273}{K}, $100.8\pm0.5\,\mathrm{(stat)}\pm0.2\,\mathrm{(laser)}\pm2.5\,\mathrm{(Sherman)}\%$ at \SI{195}{K}, and $99.4\pm0.4\,\mathrm{(stat)}\pm1.2\,\mathrm{(laser)}\pm2.5\,\mathrm{(Sherman)}\%$ at \SI{77}{K}. Values above 100\% are physical only within the calibration uncertainties.
The maxima nevertheless show that individual scans reach values statistically consistent with complete spin polarization.

\section{Surface stability measurements}
\begin{figure}[htb]
    \centering
    \includegraphics[width=0.85\columnwidth]{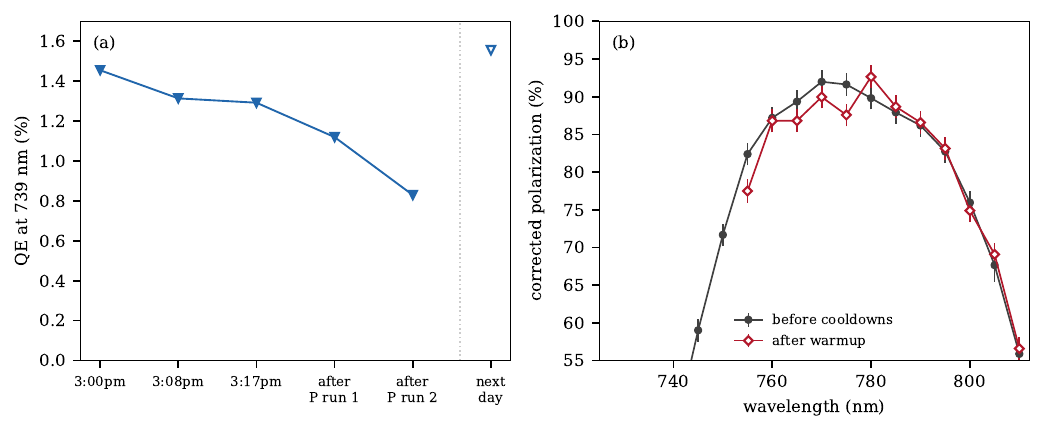}
    \caption{Surface stability measurements. (a)~QE at the \SI{77}{K} operating wavelength (\SI{739}{nm}) through the measurement day; the polarization runs took place between the marked scans, and the open point is the following day. (b)~Room-temperature polarization spectra, corrected for the laser circular polarization, before the cooldowns and after the full cooling sequence and warmup.}
    \label{fig:sm-stability}
\end{figure}

Positive-electron-affinity energy filtering is known to raise the measured polarization of bulk GaAs above the NEA value as the activation layer degrades~\cite{Levenson2025}.
The time series in Fig.~\ref{fig:sm-stability}(a) and the warmup reversibility in Fig.~\ref{fig:sm-stability}(b) bound this effect.
We reproduced the room-temperature peak polarization to $+1.1(2.1)$ percentage points after the full cooldown sequence.
At fixed temperature the QE decayed by 43\% while the polarization measurements showed no systematic increase.
A repeat dry-ice measurement on a degraded activation gave \emph{lower}, not higher, helicity-normalized polarization ($90.1(1.1)$\% versus $97.1(0.8)$\%), opposite in sign to the energy-filtering signature of Ref.~\cite{Levenson2025}.

\section{Spin-relaxation mechanisms and rates}
\label{sec:sm-rates}
Three mechanisms dominate conduction-electron spin relaxation in p-type III-V semiconductors~\cite{OpticalOrientation1984,Zutic2004,Wu2010}: D'yakonov-Perel' (DP) precession about the momentum-dependent effective field of the inversion-asymmetric lattice~\cite{Dyakonov1972}, Bir-Aronov-Pikus (BAP) electron-hole exchange scattering~\cite{Fishman1977,Zerrouati1988}, and Elliott-Yafet (EY) scattering during momentum relaxation.
Their temperature dependences differ: the DP rate grows steeply with temperature through the thermal carrier energy and the momentum-relaxation time, the BAP rate varies far more weakly, and EY is a minor contribution at our acceptor density~\cite{Chubenko2021}.
DP therefore dominates at room temperature, and BAP becomes the larger thermalized channel on cooling.

These are bulk expressions, and the active region is a superlattice.
Confinement in the \SI{3.8}{nm} wells raises the mean-squared wavevector along the growth axis well above its bulk thermal value, which modifies both the Dresselhaus splitting that drives DP and the electron-hole overlap that drives BAP.
We therefore use these mechanisms to indicate the direction of the expected temperature dependence rather than as a quantitative prediction for this structure.

Direct pump-probe measurements in a Zn-doped GaAs/GaAsP strain-compensated superlattice give $\tau_s=\SI{104}{ps}$ at room temperature~\cite{Ohki2017}.
Combined with our fitted $\tau/\tau_s(\SI{295}{K})=0.049(12)$, that value implies an effective emission time $\tau\approx\SI{5}{ps}$, a plausible transport time across the \SI{198}{nm} active region.

\section{Model fits}
\emph{Band-structure.}
We expand the superlattice period in plane waves and diagonalize a four-band Luttinger--Kohn Hamiltonian with the Bir-Pikus strain interaction for the valence minibands and a single conduction band with a Kane nonparabolicity correction, using the nominal layer structure and the recommended band parameters of Ref.~\cite{Vurgaftman2001}.
The nonparabolicity replaces the parabolic conduction excess energy $E_{\mathrm{p}}$ above the well edge by $E=[\sqrt{1+4\alpha E_{\mathrm{p}}}-1]/2\alpha$ with $\alpha=\SI{0.64}{eV^{-1}}$.
The GaAs wells are compressively strained by the relaxed GaAs$_{0.65}$P$_{0.35}$ buffer ($\varepsilon_{\parallel}=-1.26\%$), which alone splits the HH and LH edges by \SI{97}{meV}.
Confinement raises the splitting to $\delta=\SI{105}{meV}$.
For every $(k_{\parallel},q)$ point we form all valence-to-conduction transitions with the $\sigma^{+}$ matrix elements.
Each transition $i$ retains its energy $E_i$, its conduction excess energy $\Delta E_{c,i}$ above the lowest conduction state, and a polarization set by the $k_{\parallel}$-mixed eigenstates.
This step calculates the electron polarization of the strained photocathode, not only its band energies.
For transition $i$, the strained-superlattice eigenvectors and their envelope overlaps give the absorption weights $a_i^{(+)}$ and $a_i^{(-)}$ into the two conduction-electron spin projections.
We retain both the total absorption and its spin-weighted counterpart,
\begin{equation}
 a_i=a_i^{(+)}+a_i^{(-)},
 \qquad
 P_i a_i=a_i^{(+)}-a_i^{(-)},
 \label{eq:sm-transition-polarization}
\end{equation}
so that $P_i$ is the polarization created by that particular transition.
Because strain and finite-$k_{\parallel}$ mixing change the valence eigenvectors, they change $a_i^{(\pm)}$ and hence the calculated polarization as a function of photon energy.
With no fitted parameters, this calculation gives the HH edge \SI{133}{meV} above the bulk GaAs gap, $\delta=\SI{105}{meV}$, and an edge blueshift of \SI{81}{meV} from \SI{295}{K} to \SI{77}{K} (measured: \SI{78}{meV}; bulk GaAs Varshni alone: \SI{83}{meV}).
Varying the band offsets by $\pm$\SI{50}{meV} moves the edge quantities by up to \SI{12}{meV}; varying the layer thicknesses by $\pm5$\%, the barrier composition by $\pm0.02$, or the shear deformation potential by 10\% moves them by at most \SI{8}{meV} each.

\emph{Spectral model and fits.}
We assemble the calculated spin-resolved transitions into the helicity-normalized polarization spectrum
\begin{equation}
 P(\hbar\omega,T)=\frac{1}{1+\tau/\tau_s}\,
 \frac{\sum_i P_i s_i a_i g_i(\hbar\omega)+P_{\mathrm{bg}}a_{\mathrm{bg}}}
 {\sum_i a_i g_i(\hbar\omega)+a_{\mathrm{bg}}}.
 \label{eq:sm-polarization-model}
\end{equation}
Here $g_i$ broadens the calculated transition energy, $s_i$ is the spin survival during the hot-electron relaxation cascade, $a_{\mathrm{bg}}$ is an effective low-polarization contribution, and $1/(1+\tau/\tau_s)$ accounts for spin loss during transport of thermalized electrons to the surface.
We evaluate the broadening and the cascade survival as
\begin{equation}
 g_i(\hbar\omega)=\exp\!\left[-\frac{\left(\hbar\omega-E_i-\Delta E_0\right)^2}{2\Gamma^2}\right],
 \qquad
 s_i=\exp\!\left[-\left(\Delta E_{c,i}/E_\kappa\right)^{4}\right],
 \label{eq:sm-ingredients}
\end{equation}
where $\Delta E_0$ is a rigid shift of the calculated spectrum, $\Gamma$ is a fitted Gaussian width, and $E_\kappa$ sets the conduction excess energy beyond which the relaxation cascade erases the spin memory.
Coulomb attraction enters twice.
The continuum weights are multiplied by the two-dimensional Sommerfeld factor
\begin{equation}
 S(\varepsilon_i)=\frac{2}{1+\exp\left(-2\pi\sqrt{E_b/\varepsilon_i}\right)},
 \label{eq:sm-sommerfeld}
\end{equation}
where $\varepsilon_i$ is the energy of transition $i$ above its miniband-pair edge and $E_b=\SI{9}{meV}$ is the two-dimensional exciton binding energy, held fixed.
Each miniband pair within \SI{150}{meV} of the lowest edge also contributes a bound-exciton Gaussian line centered $E_b$ below its edge, with strength $f_x$ times the pair continuum weight within $E_b$ of the edge and with the polarization of that near-edge weight.
The low-polarization contribution is a broadened step at the bulk-GaAs Varshni gap $E_g(T)$,
\begin{equation}
 a_{\mathrm{bg}}(\hbar\omega)=A_{\mathrm{bg}}\,D\,\frac{1}{2}\left[1+\operatorname{erf}\!\left(\frac{\hbar\omega-E_g(T)}{\sqrt{2}\,\Gamma}\right)\right],
 \label{eq:sm-background}
\end{equation}
which carries polarization $P_{\mathrm{bg}}$; the normalization $D$ is the total superlattice absorption evaluated \SI{120}{meV} above the shifted HH edge, so $A_{\mathrm{bg}}$ is dimensionless.
The photon-energy dependence in both the numerator and denominator of Eq.~\eqref{eq:sm-polarization-model} therefore comes from the summed transitions of the strained GaAs/GaAsP superlattice; the fit does not assign a free polarization at each wavelength.

We do not assign $a_{\mathrm{bg}}$ to a unique microscopic source.
It could include photoemission from the highly doped \SI{5}{nm} GaAs surface layer.
It could also represent depth-dependent transport: near the absorption edge, the light penetrates farther into the device, so some emitted electrons may be excited deeper in the structure and undergo more spin relaxation before reaching the surface.
Because both the optical penetration depth and spin transport can change with temperature, the amplitude or polarization of such a contribution could also be temperature dependent.
The present data do not independently constrain that dependence, so the fit uses shared, temperature-independent $A_{\mathrm{bg}}$ and $P_{\mathrm{bg}}$; the extracted $\tau/\tau_s(T)$ should therefore be interpreted as an effective transport-depolarization parameter conditional on this assumption.

We fit six shared spectral-shape parameters of Eq.~\eqref{eq:sm-polarization-model} and four transport factors $\tau/\tau_s(T)$ jointly to the 48-point room-temperature spectrum and the three lower-temperature corrected peak polarizations by least squares, weighting each spectrum point by its counting-statistics error added in quadrature with a 1.5-percentage-point scan-to-scan repeatability uncertainty.
The fitted room-temperature spectrum pools all room-temperature scans, taken on separate days and activations before and after the cooldowns; the warmup comparison below shows them consistent within $+1.1(2.1)$ percentage points.
The room-temperature spectrum fixes the spectral shape, and each lower temperature enters only through its peak polarization, which determines its transport factor; each cold datum is compared with the maximum of the modeled spectrum over wavelength, weighted by the measured peak uncertainty. The Varshni laws shift the band structure between temperatures.
We fit in the corrected frame and multiply the model by the measured laser circular polarization for the delivered-frame figures.
The best-fit values are a rigid edge shift $\Delta E_0=\SI{55(2)}{meV}$, which absorbs the band-offset and layer-thickness uncertainty above; $\Gamma=\SI{32(2)}{meV}$; $E_\kappa=\SI{169(10)}{meV}$; an exciton strength $f_x=36(2)$, in units of the continuum weight within $E_b$ of each pair edge; $A_{\mathrm{bg}}=0.08(2)$; $P_{\mathrm{bg}}=0.18(14)$; and $\tau/\tau_s=0.049(12)$, $0.002(30)$, $0.000(12)$, and $0.006(17)$ at \SI{295}{K}, \SI{273}{K}, \SI{195}{K}, and \SI{77}{K}, respectively.
The parameter uncertainties come from the Jacobian at the best fit.
The fit has 51 points, 10 free parameters, and 41 degrees of freedom, giving $\chi^2_\nu=1.10$.
The 1.5-percentage-point repeatability uncertainty was set from the observed scan-to-scan scatter before fitting, and the parameter uncertainties should be read as local sensitivity estimates rather than pure counting-statistics errors.
The room-temperature spectrum alone does not separate the flat suppression $\tau/\tau_s$ from dilution by the effective low-polarization contribution.
In a fit to the room-temperature spectrum by itself, $\tau/\tau_s(\SI{295}{K})$ collapses to $0.013(14)$ and the background absorbs the suppression.
Profiling the joint fit over $\tau/\tau_s(\SI{295}{K})$ places the minimum between 0.04 and 0.06 and disfavors zero by $\Delta\chi^{2}=42$, so the quoted value is a joint constraint.
The fitted decrease is much faster than any thermalized spin-relaxation rate produces. $\tau/\tau_s$ falls from its room-temperature value to a value consistent with zero by \SI{273}{K}, a \SI{22}{K} interval over which the bulk mechanisms of Sec.~\ref{sec:sm-rates} change only marginally. Because $\tau/\tau_s$ is an effective parameter conditional on the fixed-background assumption, this mismatch may reflect temperature dependence of the emission time $\tau$, relaxation channels beyond the thermalized-rate picture, and/or the low-polarization contribution itself.

\begin{figure}[htb]
    \centering
    \includegraphics[width=\columnwidth]{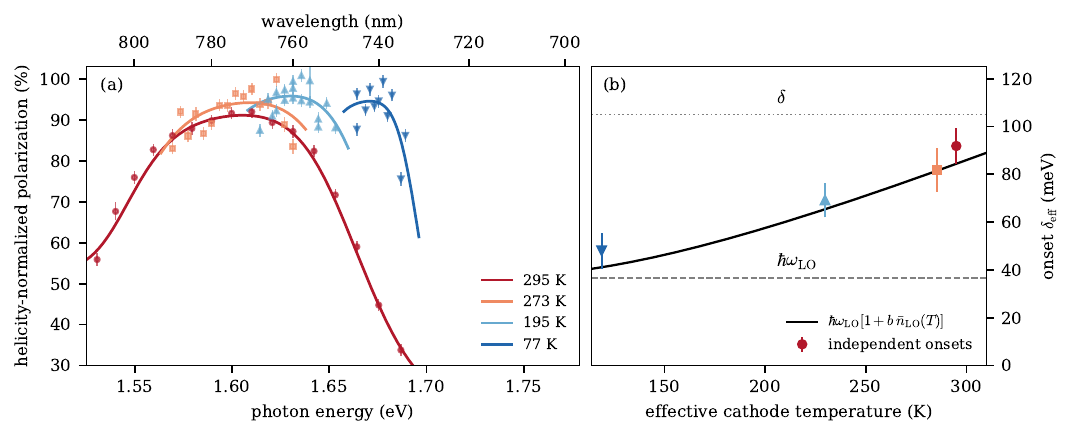}
    \caption{Phenomenological description of the peak narrowing. (a)~Helicity-normalized polarization spectra at the four coolant settings, with the simplified two-edge description using the LO-occupation onset law. Photon energy is used on the bottom axis because the narrowing and onset are quantified as energy intervals; the corresponding wavelength is shown on the top axis for comparison with the main-text spectra. (b)~Blue-side depolarization onset $\delta_{\mathrm{eff}}$ extracted independently at each setting (symbols) versus the peak-shift-inferred cathode temperature, with the one-parameter law $\hbar\omega_{\mathrm{LO}}[1+b\,\bar{n}_{\mathrm{LO}}(T)]$, $b=4.2(5)$ (solid line), shown across the full displayed temperature range. Dashed line: the LO-phonon energy, the $T\to0$ limit of the law; dotted line: the band-structure HH--LH splitting $\delta=\SI{105}{meV}$, which the onset approaches at room temperature but falls far below on cooling.}
    \label{fig:blue-edge}
\end{figure}

\emph{Peak widths below room temperature.}
Equation~\eqref{eq:sm-polarization-model} with room-temperature parameters does not reproduce the narrowing of the polarization peak on cooling, and we do not fit the widths with it.
We instead use a simplified two-edge description of the narrowing: a heavy-hole step, an oppositely polarized light-hole step at an effective splitting $\delta_{\mathrm{eff}}(T)$ above it, a weakly polarized background, and transport factors fixed by the cold-saturation analysis,
\begin{equation}
 P(\hbar\omega,T)=\frac{1}{1+r_T}\,
 \frac{3\Phi_{\mathrm{HH}}-A_{\mathrm{LH}}\Phi_{\mathrm{LH}}+P'_{\mathrm{bg}}A'_{\mathrm{bg}}\Phi_{g}}
      {3\Phi_{\mathrm{HH}}+A_{\mathrm{LH}}\Phi_{\mathrm{LH}}+A'_{\mathrm{bg}}\Phi_{g}},
 \label{eq:sm-two-edge}
\end{equation}
where each $\Phi_X=\frac{1}{2}\{1+\operatorname{erf}[(\hbar\omega-E_X)/\sqrt{2}\,\Gamma_T]\}$ is a Gaussian-broadened step with width $\Gamma_T=\Gamma_0+c\,k_BT_{\mathrm{eff}}$, the heavy-hole edge sits at $E_{\mathrm{HH}}=E_g(T_{\mathrm{eff}})+E_{\mathrm{off}}$, the light-hole edge at $E_{\mathrm{LH}}=E_{\mathrm{HH}}+\delta_{\mathrm{eff}}(T)$, and the background edge at the bulk gap $E_g(T_{\mathrm{eff}})$.
The factor of 3 is the bulk $\sigma^{+}$ selection-rule weight of the heavy-hole channel, and the primed amplitudes are fitted within this parameterization, distinct from the parameters of Eq.~\eqref{eq:sm-polarization-model}.
The fitted thermal component of the width is consistent with thermal occupation of the initial states at our doping.
The effective cathode temperature $T_{\mathrm{eff}}$ comes from the measured peak shift.
We locate each measured peak energy by a local quadratic fit to the helicity-normalized points within \SI{35}{meV} of the maximum, and we form a peak thermometer by evaluating the peak energy of the fitted Eq.~\eqref{eq:sm-polarization-model} spectrum at the four nominal temperatures and interpolating monotonically between them.
The \SI{295}{K} peak supplies one absolute energy anchor, and inverting the thermometer maps each colder peak position to $T_{\mathrm{eff}}=\SI{286}{K}$, \SI{230}{K}, and \SI{119}{K} at the \SI{273}{K}, \SI{195}{K}, and \SI{77}{K} coolant settings.
The transport factor of this description, $r_T$, plays the role that $\tau/\tau_s$ plays in Eq.~\eqref{eq:sm-polarization-model}, and it is fixed before the fit rather than fitted. The weighted mean of the statistically consistent \SI{195}{K} and \SI{77}{K} helicity-normalized peaks defines a transport-free ceiling $P_\infty$, and $r_T=\max\left[0,\,P_\infty/P_{\mathrm{peak}}(T)-1\right]$.
The \SI{77}{K} scans cover \SIrange{734}{745}{nm}, and we infer the width there from this description.
Extracted independently at each temperature, the onset is $\delta_{\mathrm{eff}}=\SI{92(8)}{meV}$, \SI{82(9)}{meV}, \SI{69(7)}{meV}, and \SI{48(8)}{meV} at the \SI{295}{K}, \SI{273}{K}, \SI{195}{K}, and \SI{77}{K} coolant settings [Fig.~\ref{fig:blue-edge}(b)].
At room temperature the onset is close to the band-structure HH--LH splitting $\delta=\SI{105}{meV}$. At the \SI{77}{K} coolant setting it falls far below $\delta$, so an energy-dependent depolarization process, not the band structure, sets the low-temperature blue side.
The one-parameter law $\delta_{\mathrm{eff}}(T)=\hbar\omega_{\mathrm{LO}}\,[1+b\,\bar{n}_{\mathrm{LO}}(T)]$, with $\hbar\omega_{\mathrm{LO}}=\SI{36.7}{meV}$ the GaAs LO-phonon energy, $\bar{n}_{\mathrm{LO}}$ its Bose occupation at the inferred cathode temperature, and $b=4.2(5)$ fitted, provides a compact empirical description of all four spectra [Fig.~\ref{fig:blue-edge}(a)].
The onset grows with the phonon occupation and approaches the single LO-phonon energy as the phonons freeze out.
This correlation is consistent with LO-phonon freeze-out contributing to the narrowing, but the law is not derived from a microscopic model and does not identify LO phonons as the unique cause.

\clearpage

\section{Benchmark comparison}
Table~\ref{tab:sm-benchmark} collects published peak spin polarizations and quantum efficiencies from GaAs-family and alkali-antimonide photocathodes, ordered chronologically.
The cited papers define and measure QE at different wavelengths and operating points, so the QE column supports qualitative comparison only.
\begin{table}[!ht]
\caption{Published peak spin polarization and QE from spin-polarized photocathodes, ordered chronologically. QE values are reported as given in the cited papers and may refer to peak QE or QE at the polarization-optimized wavelength; comparisons of QE should therefore be interpreted qualitatively. For the MOCVD-grown DBR device of Belfore \emph{et al.}, the listed polarization and QE were measured together at \SI{785}{nm}, corresponding to $P^2\mathrm{QE}=1.95\%$. Central polarization values are listed; differences between rows are not necessarily statistically significant given the quoted uncertainties. The polarizations for this work are helicity-normalized using the measured circular polarization of the excitation; their uncertainties are statistical plus scan-to-scan scatter, laser calibration, and the Sherman function. The delivered beam polarizations, measured with 97.8\% (\SI{195}{K}) and 96.3\% (\SI{77}{K}) circularly polarized light, are $95.0(0.8)$\% and $90.8(1.3)$\%.}
\label{tab:sm-benchmark}
\begin{ruledtabular}
\begin{tabular*}{\textwidth}{@{\extracolsep{\fill}}llllr}
Photocathode & Peak $P_{\mathrm{e}}$ & QE & Year & Reference \\
\colrule
Bulk GaAs NEA & $\sim$40\% & -- & 1975/1976 & \cite{Pierce1975,Pierce1976} \\
Strained InGaAs & $\sim$71\% & -- & 1991 & \cite{Maruyama1991} \\
Strained GaAs epilayer & $86\pm10$\% & $2.4\times10^{-4}$ & 1991 & \cite{Nakanishi1991} \\
AlGaAs--GaAs SL & $71.2\pm1.1\pm6.1$\% & $2.7\times10^{-6}$& 1991 & \cite{Omori1991} \\ %
GaAs/GaAsP SL & 86\% & 1.2\% & 2004 & \cite{Maruyama2004} \\
GaAs/GaAsP, InGaAs/AlGaAs SL & $92\pm6$\% & 0.5\% & 2005 & \cite{Nishitani2005} \\
InAlGaAs/GaAs SL & 91\% & 0.14\% & 2006 & \cite{Gerchikov2006} \\
AlInGaAs/AlGaAs SL & 92\% & 0.85\% & 2008 & \cite{Mamaev2008} \\
GaAs/GaAsP compensated SL & 92\% & 1.6\% & 2013 & \cite{Jin2013} \\
DBR GaAs/GaAsP SL & 84\% & 6.4\% & 2016 & \cite{Liu2016} \\
Alkali antimonide (Na$_2$KSb/Cs$_3$Sb) & 40--50\% & $\lesssim$15\% & 2022 & \cite{Rusetsky2022} \\
DBR GaAs/GaAsP compensated SL & $>$75\% & $\sim$10\% & 2023 & \cite{Biswas2023} \\
DBR GaAs/GaAsP SL (MOCVD) & 82\% & 2.9\% & 2023 & \cite{Belfore2023} \\
Cryogenic GaAs/GaAsP SL (\SI{195}{K}) & $97.1\pm0.8\pm0.3\pm2.4$\% & 0.7\% & 2026 & this work \\
Cryogenic GaAs/GaAsP SL (\SI{77}{K}) & $94.3\pm1.3\pm1.2\pm2.3$\% & 1.3\% & 2026 & this work \\
\end{tabular*}
\end{ruledtabular}
\end{table}

\bibliography{references}